# Real-time Calibration-free Imaging Through Dynamic and Distinct Multimode Fibers via Spatial Harmonic Invariant Nonlinear Encoding (SHINE)

Zhiyuan Wang[1,†], Haoran Li[1,†], Songjie Luo[2], Jixiang Chen[3], Tianting Zhong[3,4], Jing Yao[1,4], Jixiong Pu[2], Zhipeng Yu[1,4,*], Sylvain Gigan[5,*], Ziyang Chen[2,*], and Puxiang Lai[1,4,6,*]

[1] Department of Biomedical Engineering, The Hong Kong Polytechnic University, Hong Kong, China
[2] College of Information Science and Engineering, Huaqiao University, Fujian Provincial Key Laboratory of Light Propagation and Transformation, Xiamen, China, 361021
[3] Department of Electrical and Electronic Engineering, The University of Hong Kong, Hong Kong, China
[4] Shenzhen Research Institute, The Hong Kong Polytechnic University, Shenzhen, China
[5] Laboratoire Kastler Brossel, ENS-Université PSL, CNRS, Sorbonne Université, Collège de France, Paris, France
[6] Photonics Research Institute, The Hong Kong Polytechnic University, Hong Kong, China

**Abstract:** Multimode fibers (MMFs) provide a compact, high-throughput platform for minimally invasive imaging and information transmission. However, their utility is fundamentally constrained by mode mixing, which renders image transmission spatially disrupted and sensitive to external perturbations. Current imaging methods typically rely on transmission matrix measurement or deep learning models that are fragile to fiber movement, necessitating frequent, time-consuming calibrations and re-calibrations that are easily disrupted and fail to generalize across different fiber configurations—let alone across entirely distinct fibers. Here, we propose a calibration and feedback-free MMF coherent imaging paradigm, that we termed Spatial Harmonic Invariant Nonlinear Encoding (SHINE). By leveraging the angle-dependent phase-matching conditions of second-harmonic generation, we encode spatial features into broadband spectral signatures that possess intrinsic insensitivity not only to modal scrambling but also to fiber bending, movement, as well as structural variations. This spectral representation enables a deep learning model to robustly reconstruct images in real time despite dynamic perturbations and even generalizes well to distinct MMFs without recalibration or feedback. We achieve experimentally an average Pearson correlation coefficient (PCC) of 0.82 for image reconstruction tasks on Fashion-MNIST and a classification accuracy of 92.3% on HERLEV biomedical dataset. Uniquely, our method exhibits remarkable cross-fiber generalization: a model trained on a single MMF successfully reconstructs images transmitted through entirely distinct, previously unseen MMFs with a PCC of 0.74. These results establish a robust, calibration-free framework for imaging through MMFs in real time, paving the way for practical, resilient optical diagnostics that operate without distal-end feedback.

*Address correspondence to Zhipeng Yu at yu.zh.yu@polyu.edu.hk, Sylvain Gigan at sylvain.gigan@lkb.ens.fr, Ziyang Chen at ziyang@hqu.edu.cn, and Puxiang Lai at puxiang.lai@polyu.edu.hk

†These authors contributed equally to this work.

## Introduction

Multimode optical fibers (MMFs), with diameters comparable to a human hair, support thousands of spatial modes and offer compelling advantages for high-capacity optical communication, minimally invasive imaging, and remote sensing[1-3]. However, their utility is fundamentally constrained by modal dispersion and intermodal coupling, which scramble the spatial structure of incident light[4-6]. Consequently, coherent signals entering the proximal end are transformed into seemingly stochastic speckle patterns at the distal end[7-9], disrupting information or image transmission.

To recover information from these scrambled outputs, various strategies have been explored, ranging from iterative wavefront shaping[10,11], transmission matrix (TM) calibration[2-4,12] to deep learning (DL)-based methods[13-15]. While effective under static conditions, these techniques often fail when the fiber undergoes dynamic perturbations[12,15,16], such as bending, twisting, or thermal fluctuations, or when a distinct MMF is employed. To mitigate these limitations, prior efforts have focused on periodic recalibration or dynamic tracking of the fiber's state. These methods typically rely on a distal beacon to monitor the fiber's configuration, updating the transmission mapping via digital phase conjugation[17] or TM remeasurement[2,18-20]. However, such remedies require recurrent recalibration (often frame-by-frame)[8,17,18] or extensive database matching[2], entailing high-dimensional matrix inversions, continuous bidirectional feedback, significant data storage, and even interruptions to online imaging. Alternatively, "blind" DL-based methods[5,7,21-23] attempt to bypass explicit TM inversion by training neural networks on massive datasets encompassing various deformation states. Yet, due to the high sensitivity of MMFs to perturbations, only a limited subset of principal modes remains invariant under deformation[24,25]. Consequently, even with exhaustive data acquisition, existing methods are typically effective only for relatively simple objects and struggle to generalize under severe perturbations[5,7], let alone across different MMFs. Ultimately, whether performed explicitly or embedded implicitly within training data, recalibration remains an unavoidable, high-overhead compromise.

A path towards resolving this bottleneck lies in identifying light properties that remain invariant during propagation through dynamic and distinct MMFs. Unlike conventional spatial parameters (e.g., amplitude, phase, and polarization), which are randomized by multimode interference, the optical power spectrum, representing the wavelength-dependent power distribution, remains conserved under linear propagation. This spectral invariance suggests a route for robust information transmission known as spatial-spectral encoding (SSE)[26-28]. However, translating SSE into a practical imaging modality has been hindered by the inherent limitations of linear encoders. Whether utilizing random scattering media[27] or structured metasurfaces[28], these linear approaches face a strict trade-off between spectral distinguishability and bandwidth. Such mappings are often ill-posed, suffering from severe spectral overlaps that necessitate complex sparse reconstruction algorithms. Consequently, despite the use of sophisticated nanofabrication, the achievable spatial resolution remains severely restricted, typically saturating at a few tens of pixels. This inability to scale to high-fidelity imaging renders linear SSE insufficient for practical applications.

To transcend these intrinsic limitations of linearity, we draw inspiration from second-harmonic generation (SHG), a $\chi^{(2)}$ nonlinear process governed by strict, angle-dependent phase-matching conditions. Unlike linear spatio-spectral encoders, where pixel-to-wavelength correspondence is often rigid or ill-posed, the non-colinear interaction in a nonlinear crystal acts as a deterministic angular filter. This process coherently maps the object's spatial frequency components (angular spectrum) into distinct spectral slices of the generated second-harmonic wave[29-31]. Crucially, by relying on bulk phase-matching rather than positional interference or micro-resonances, this approach inherently avoids the instability of scattering media and the nanofabrication requirements of metasurfaces, providing a physically robust foundation for encoding information.

Building upon this principle, we introduce a strategy termed Spatial Harmonic

Invariant Nonlinear Encoding (SHINE). This approach leverages the sensitivity of nonlinear optics to encode spatial information and the spectral invariance of MMFs to transmit it. As shown in Fig. 1a, in SHINE, the object's spatial information is carried by broadband ultrashort pulses into a nonlinear crystal, generating distinct SHG spectra imprinted with the object's spatial features. These spectral signatures remain invariant under MMF perturbations, unaffected by modal scrambling. This allows a single calibration step to establish a robust mapping between object patterns and output spectra, eliminating the need for repeated recalibrations or iterative wavefront corrections. To exploit this deterministic yet highly complex mapping, where spatial information is distributed across high-dimensional spectral features, we developed a deep neural network, the SHG Spectrum Decoding U-net (SSDU, see *Supplementary Note 2*), trained on a dataset of known object patterns and their corresponding nonlinear spectra. Once trained, the SSDU learns the complex inverse mapping from spectral signatures back to spatial images, allowing for real-time, robust object reconstruction and classification using spectral measurements alone, even when switching between different MMFs.

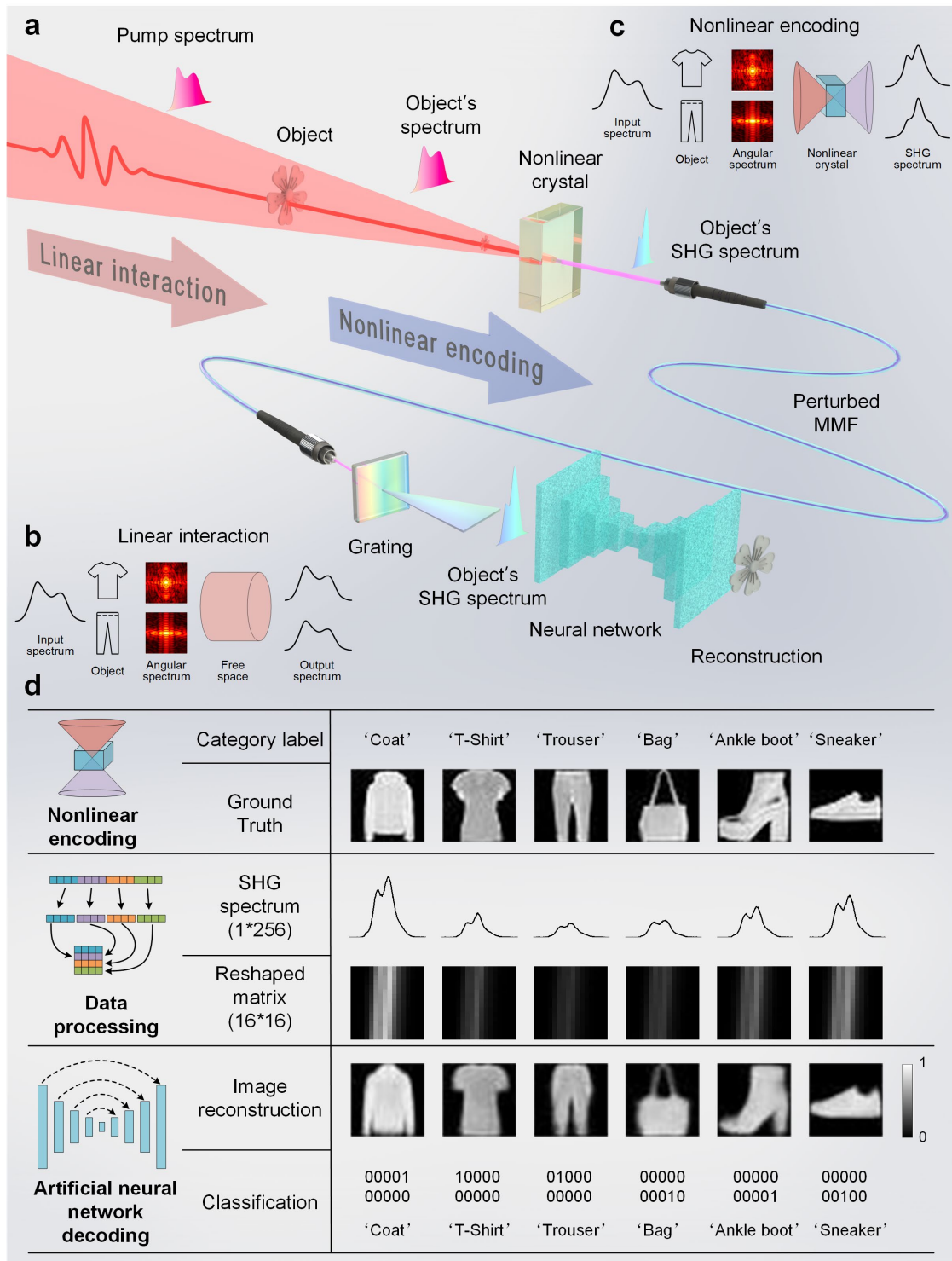

**Fig. 1 Principle and workflow of Spatial Harmonic Invariant Nonlinear Encoding (SHINE) for imaging through perturbed MMFs. a,** SHINE encodes spatial information into nonlinear spectral signatures via SHG in a crystal illuminated by a broadband pump. **b–c,** With linear interaction in free space (Inset **b**), the object's angular spectrum propagates without spatial-spectral coupling, yielding highly correlated responses; with nonlinear interaction (Inset **c**), it maps the angular spectrum into unique

SHG spectra driven by angle-dependent phase-matching conditions. The generated spectral signatures remain invariant to modal scrambling, enabling robust imaging through dynamically perturbed or entirely distinct fibers without recalibration. **d**, Schematic of the reconstruction pipeline. Spatially modulated light generates object-specific SHG spectra, which are recorded (256 points), reshaped into 16 × 16 matrices, and fed into a deep neural network for simultaneous image reconstruction and classification.

We experimentally validate the proposed SHINE method through a series of demanding tests, including single calibration imaging under dynamic bending, shaking, and twisting and across different MMFs. We demonstrate that a single calibration, performed initially under static conditions, is sufficient for accurate image reconstruction even as the fiber undergoes significant deformation or is replaced with a distinct fiber. Remarkably, the entire imaging process, from illumination to spectral acquisition and neural reconstruction, operates without distal-end feedback or additional calibration. This fundamentally differentiates SHINE from existing methods, establishing a novel hybrid paradigm that marries nonlinear optics with machine learning to pave the way towards robust, real-time, minimally invasive imaging deep within biological tissues.

## Results

### Theoretical framework: spatial-to-nonlinear spectral mapping

The core principle of SHINE lies in mapping spatial complexity into spectral diversity via angle-dependent phase matching. A broadband ultrashort pulse illuminates an object, imparting a spatially varying phase that produces a complex field at the object plane: $E_{\text{obj}}(x,y,\omega) = E_0(x,y,\omega)e^{i\phi_{\text{obj}}(x,y)}$. A Fourier lens (focal length $f$) transforms this field, yielding an angular spectrum distribution at the nonlinear crystal's entrance face:

$$A(\omega,\theta_x,\theta_y) \propto \iint E_0(x,y,\omega)e^{i\phi_{\text{obj}}(x,y)}e^{-i\frac{k(\omega)}{2f}(x^2+y^2)}e^{-ik(\omega)(x\theta_x+y\theta_y)}dxdy, \qquad (1)$$

with $\theta_x$ and $\theta_y$ the propagation angles relative to the optical axis.

Inside a $\chi^{(2)}$ nonlinear crystal, SHG is governed by strict phase-matching conditions. For an ideal plane wave under perfect phase matching, SHG emerges as a narrow beam with high efficiency. When the incident wavefront is spatially structured by phase masks, scattering, or spatial light modulators, the field must instead be described as a superposition of angular components, each with distinct transverse wavevectors[31-33]. These components propagate at varying angles inside the crystal, experiencing angle-dependent phase matching and accumulating distinct phase mismatches $\Delta k(\omega, \alpha(\theta_x, \theta_y))$. Consequently, the SHG efficiency for each angular component is modulated by a sinc-squared envelope, $\text{sinc}^2\left[\frac{L}{2}\Delta k(\omega, \alpha(\theta_x, \theta_y))\right]$ [34,35], where $L$ is the crystal length. Thus, the spatial profile of the input field shapes the angular distribution of SHG efficiency[36].

Under the undepleted pump approximation, the total SHG signal is the integral over all angular components, giving the spectral formulation[35,37]:

$$S(2\omega) \propto \iint d\theta_x d\theta_y \left|A(\omega, \theta_x, \theta_y)\right|^2 \text{sinc}^2\left(\frac{\Delta k(\omega, \alpha(\theta_x, \theta_y))L}{2}\right), \qquad (2)$$

where $A(\omega, \theta_x, \theta_y)$ is the input-field angular spectrum. This formulation reveals that both amplitude and phase of the input spatial profiles directly reshape the SHG power spectrum. Leveraging this intrinsic angular sensitivity, spatial information is effectively "imprinted" onto the spectral domain, creating robust signatures suitable for transmission through disordered media.

**Single-calibration imaging and classification under dynamic perturbations**

To demonstrate the system's resilience, gray-scale images from the Fashion-MNIST[38] dataset (10 categories) were encoded into SHG spectra. As shown in Fig. 1d, the object-modulated light was focused into the nonlinear crystal, and the resulting SHG spectral were recorded with 256 spectrometer pixels. For efficient neural network processing, the spectra data (256 points) were reshaped into 16 × 16 matrices.

The dataset collected in static condition was randomly split into training, validation, and test sets in a ratio of 8:1:1. During dynamic testing, the MMF was deliberately coiled into multiple loops and subjected to random shaking and crimping to emulate rapid and severe environmental perturbations encountered in practical scenarios (see Supplementary Video). The dynamic test set consisted of 10,000 previously unseen images collected two hours after training. Importantly, the SSDU neural network was trained exclusively using data acquired under static fiber conditions.

Figure 2a summarizes the quantitative performance under both conditions. Despite significant mechanical perturbations applied to the MMF during testing, the reconstruction quality remains largely unaffected. The Pearson Correlation Coefficient (PCC) for image reconstruction reaches 0.81 under dynamic perturbations, decreasing by only ~0.01 compared to the static case. Structural Similarity Index (SSIM) and Peak Signal-to-Noise Ratio (PSNR) follow similar trends, experiencing merely ~1.7% reduction from static to dynamic fiber condition. Similarly, the recognition accuracy stays above 70%, indicating strong robustness at both pixel and semantic levels.

Despite the limited entropy of the spectra, the system demonstrates excellent classification capabilities. Fig. 2b illustrates the confusion matrix; the network maintains 80–90% accuracy for most categories, with primary errors occurring only between visually similar classes, such as T-shirts (0), pullovers (2), coats (4), and shirts (6), which differ only by subtle features like collars and sleeves.

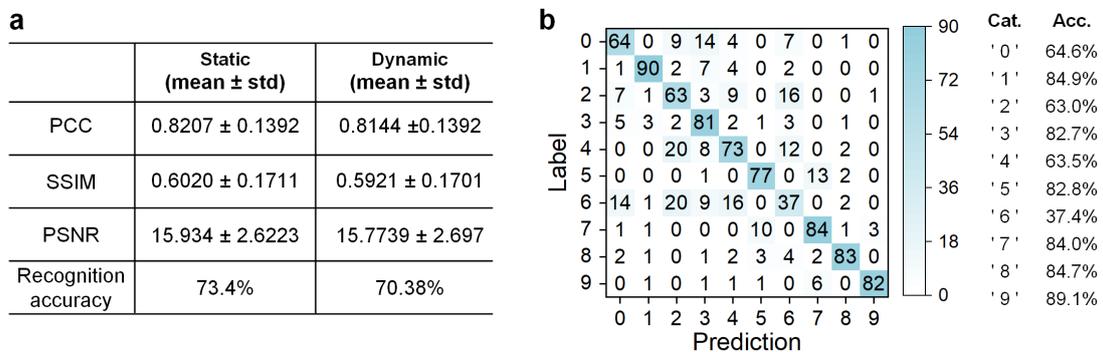

**Fig. 2 Performance evaluation on the Fashion-MNIST dataset. a,** Quantitative evaluation metrics (PCC, SSIM, and classification accuracy) comparing performance in a static fiber versus a dynamically

perturbed fiber. **b,** Confusion matrix for classification on the test set. Class labels: 0-T-shirt, 1-Trouser, 2-Pullover, 3-Dress, 4-Coat, 5-Sandal, 6-Shirt, 7-Sneaker, 8-Bag, 9-Ankle boot.

Finally, the neural network successfully bridges the underdetermined inversion (256→784 pixels) by combining physically invariant encoding with data-driven decoding, enabling faithful recovery of the most relevant spatial features. Notably, the neural network, trained in static conditions, demonstrates strong generalization ability to dynamic perturbed regimes, due to its primary reliance on spectral amplitude features rather than spatial interference patterns. This confirms that SHINE converts perturbation-sensitive spatial information into stable spectral fingerprints, enabling reliable imaging without recalibration.

**Biomedical imaging and classification on the HERLEV dataset**

To evaluate the method's potential for biomedical applications, we applied SHINE to a more complex dataset —HERLEV[39], which consists of 917 cervical cell images across seven diagnostic categories. All images were resized to 224×224 pixels. From each category, 10 images were randomly selected for testing (70 in total). The remaining images were split for training (800) and validation (47). To augment the training data, each image underwent nine transformations (flipping, rotation, blurring, contrast enhancement, brightness adjustment, and magnification), resulting in 8,000 training, 470 validation, and 700 testing samples to enhance model robustness.

Fig. 3a presents representative reconstructions. Despite the high compression ratio (reconstructing 224 × 224 pixel images from just 256 spectral points), the network faithfully reproduces global cell morphology, including the cytoplasmic boundaries and nuclear intensity distributions. Fig. 3b shows reconstruction results for augmented samples and highlights the network's sensitivity to fine spectral features. The figure also demonstrates the reconstruction robustness under geometric transformations, particularly rotation. Theoretically, due to birefringent phase matching, the angular acceptance is generally anisotropic along the principal axes. However, in our configuration, we did not observe a pronounced direction-dependent reconstruction bias at the image level, likely because the effective angular support was limited by the collection numerical aperture (NA) and the network learns from the full mixed spectral signature. Nevertheless, anisotropic acceptance could be exploited deliberately in future designs to tailor spectral sensitivity.

To interpret this decoding capability physically, we analyzed the spectral response to spatial transformations (Fig. 3d–e). As shown in Fig. 3d, although rotational transformations (Rot-1 to Rot-3), brightness weighting, and scaling operations generate spectra that are visually highly similar to the original one and still exhibit high Pearson correlation coefficients (>0.99), the differences introduced by these operations can be more clearly captured by the Euclidean distance. The Euclidean distance exhibits distinct non-zero values, indicating that orientation information is encoded in subtle local intensity shifts rather than global shape deformation. Conversely, weighting induces massive deviations in Euclidean distance, reflecting macroscopic changes in total energy. This separation is further visualized in the principal component analysis (PCA) latent space (Fig. 3e). Geometric rotations and scale cluster locally near the original sample but remain separable along PC2, whereas photometric changes (weight) follow distinct trajectories along PC1. This confirms that the SHG spectrum encodes geometric and photometric variations in orthogonal feature subspaces, which the neural network successfully disentangles.

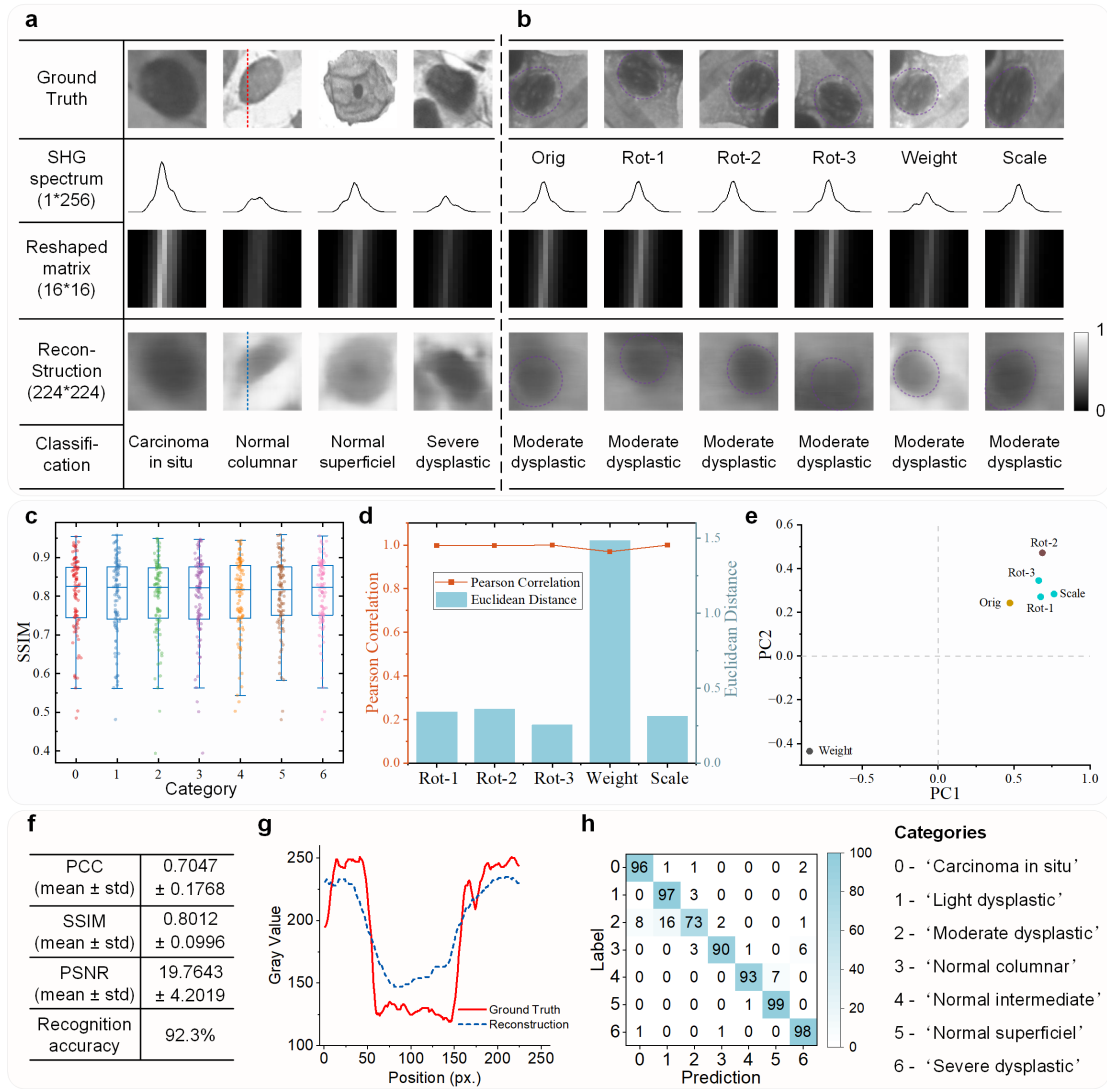

**Fig. 3 Reconstruction and classification results on the HERLEV dataset. a,** Representative examples showing the reconstruction workflow and classification outcomes. From top to bottom: ground-truth cell images, measured SHG spectra, reshaped spectral matrices (16 × 16), reconstructed images (224 × 224), and predicted diagnostic categories. **b,** Reconstruction robustness under data augmentation. The dashed region highlights transformed inputs: including (from left to right) original, rotations of 90°, 180°, and 270°, weighted addition, and magnification. **c,** Distribution of Structural Similarity Index (SSIM) across the seven diagnostic categories. **d,** Quantitative sensitivity analysis of SHG spectra to spatial transformations. The line plot (left axis) shows Pearson Correlation, indicating shape similarity to the original input, while the bar chart (right axis) shows Euclidean Distance, quantifying absolute spectral deviation. **e,** PCA projection of the spectral dataset, visualizing the distinct trajectories of geometric (rotation) versus photometric (weight/scale) transformations in the latent feature space. **f,** Statistical

evaluation metrics (PCC, SSIM) and overall recognition accuracy on the test set. **g**, Cross-sectional intensity profiles comparison reconstructed and ground-truth images along the cross-sections indicated by the red and blue dashed lines in **a**. **h**, Confusion matrix for diagnostic classification on the HERLEV test set.

Consequently, the reconstruction quality remains robust across all diagnostic categories. The boxplot distribution of SSIM (Fig. 3c) shows consistent performance across the seven classes, with median values stably around 0.80, indicating that the encoding efficacy is not biased by specific pathological morphologies. Overall, the system achieves a mean PCC of 0.70 ± 0.18 (Fig. 3f) and strong agreement in cross-sectional intensity profiles (Fig. 3g). Most notably, the overall recognition accuracy on the HERLEV test set reaches 92.3%. Misclassifications are primarily confined to adjacent pathological grades (e.g., "light" vs. "moderate" dysplasia), which share highly similar morphological—and thus spectral—features in real specimens (Fig. 3h).

These results collectively demonstrate that the SHG spectrum captures a compact yet diagnostically rich representation of biological structures. The low- and mid-frequency components in the angular spectrum determine the morphological outlines, while subtle anisotropies and intensity variations encode positional and brightness details. Through data-driven learning, the network establishes a strong prior mapping between spectral signatures and spatial patterns, effectively bridging the underdetermined reconstruction problem. This synergy between the physical encoding of SHG spectra and the statistical prior of deep learning enables remote cytology through perturbed MMFs.

**Calibration-free imaging and classification across distinct MMFs**

Beyond resilience to dynamic perturbations, we evaluated the generalizability of SHINE across distinct MMFs with varying geometries and manufacturers. As detailed in Fig. 4a, we tested five MMFs: A1 and A2 (Thorlabs M15L10, core diameter: 105 μm, length: 10 m, NA = 0.22), B1 (Xinray SUH105, 105 μm, 2 m, NA = 0.22), B2 (Xinray SUH105, 105 μm, 10 m, NA = 0.22), and B3 (Xinray SUH200, 200 μm, 2 m, NA = 0.22). A1 and A2 are of identical model from the same manufacturer (Thorlabs), while

B1-B3 are sourced from Xinray but differ in both length and core diameter. Crucially, each MMF was connected to the spectrometer independently, without precise alignment or configuration optimization, emulating realistic, unconstrained "plug-and-play" deployment. For each configuration, 10,000 Fashion-MNIST images were encoded and recorded, split into standard training, validation, and test sets with a ratio of 8:1:1. We trained a model exclusively on dataset from one MMF and then evaluated it on the test sets of all fibers without any fine tuning or calibration.

Fig. 4b shows the cross-evaluation matrix. As seen, reconstruction performance between A1 and A2, the identical-model fibers from the same manufacturer, shows minimal deviation, confirming intrinsic reproducibility. Notably, models trained on A1 and A2 generalize robustly to B2 and B3, maintaining PCC values above 0.70. This indicates that the system is surprisingly resilient to variations in fiber manufacturer and even core diameter (as seen with the 200 μm B3 fiber). However, a noticeable performance drop is observed when generalizing to or from B1, with PCC values decreasing to the 0.55–0.64 range.

To elucidate the physical origins of the generalization disparity, particularly the poor transferability observed in fiber B1, we conducted a detailed comparative analysis of the spectral characteristics across all fibers (Fig. 4d–f). Fig. 4d presents the spectral consistency; while the macroscopic envelopes of the representative sample (upper panel) and class-averaged spectra (lower panel) share similar spectral profiles despite intensity variations caused by coupling efficiency or fiber attenuation, B1 exhibits noticeable deviations from the trends established by the other fibers. This spectral distinctiveness is more rigorously quantified in the frequency domain analysis (Fig. 4e), which compares the average magnitude of the Fast Fourier Transform (FFT) of the spectra. Notably, fibers A1 and A2 (blue and red curves) display nearly identical distributions, with B2 (purple), which shares the same 105 μm core and 10 m length, closely following this pattern. In stark contrast, B1 (yellow) emerges as a significant outlier with a divergent spectral modulation frequency distribution, followed by B3 (green). Furthermore, spectral information entropy analysis (Fig. 4f) reveals that B1 possesses the lowest information capacity. Collectively, these results suggest that the specific geometry of B1 (short length combined with a small core) results in insufficient modal mixing. Notably, comparison between B1 and B3 offers deeper insight into the physical threshold for generalization. While B1 (105 µm, 2 m) suffers from severe domain shift, B3 (200 µm, 2 m) maintains high generalization performance despite sharing the same short length. This indicates that the larger core diameter of B3 supports a significantly higher number of guided modes, which effectively compensates for the reduced propagation length. Consequently, B3 achieves a level of modal scrambling and information entropy comparable to the 10 m fibers (A1, A2, B2), whereas B1 falls below the critical threshold for adequate modal mixing.

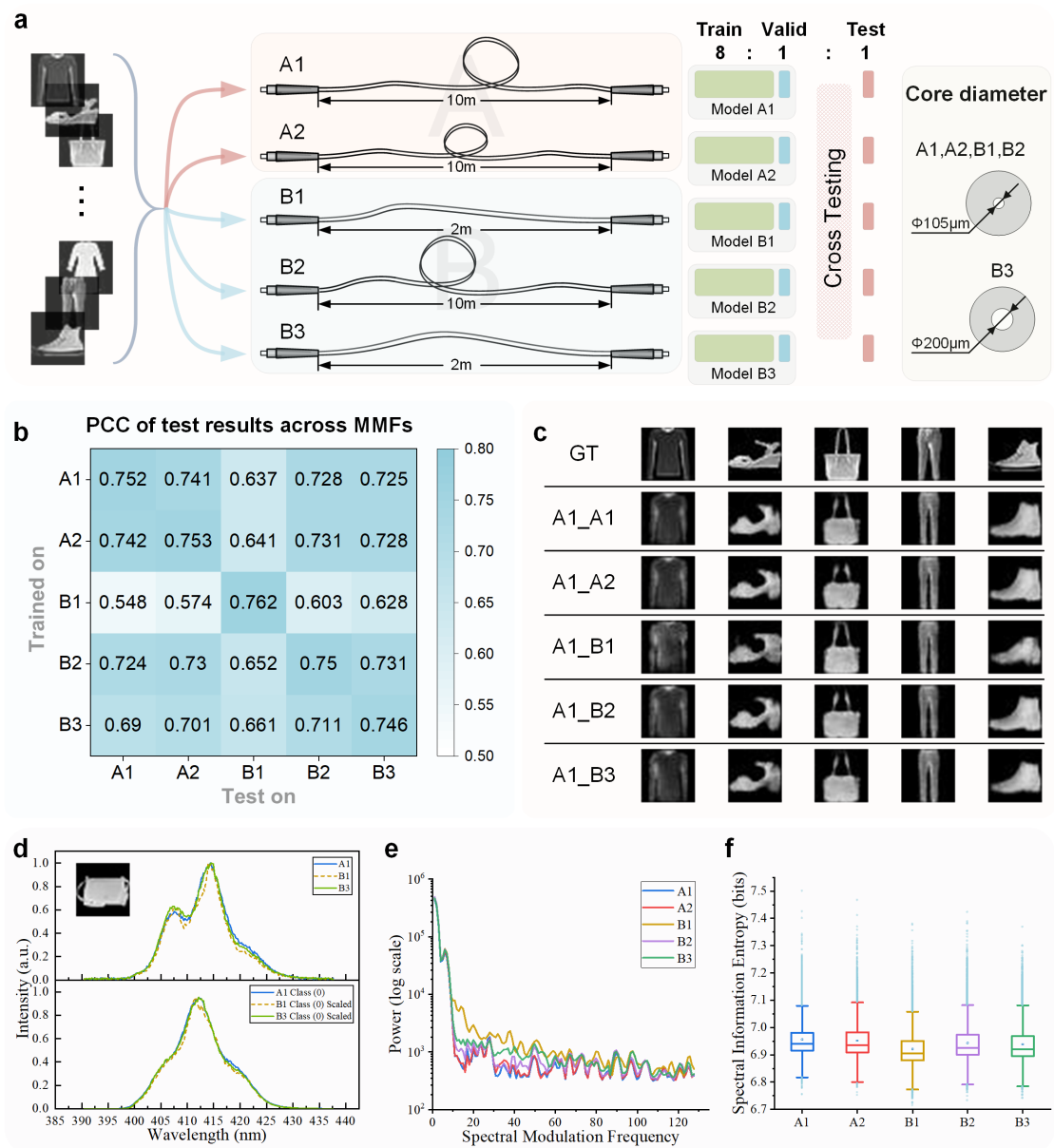

**Fig. 4 Cross-fiber generalization capabilities. a**, Experimental strategy for assessing generalization across five distinct MMFs. Models were trained exclusively on data from a single fiber (e.g., Fiber A1) and tested on the trained fiber and unseen distinct fibers (e.g., Fibers A2–B3). **b**, Cross-evaluation matrix showing the average Pearson Correlation Coefficient (PCC) for image reconstruction. Rows represent the fiber used for training, while columns represent the fiber used for testing. **c**, Representative reconstruction results using a model trained on fiber A1 and tested on datasets from all five distinct fibers (A1–B3). **d**, Spectral intensity comparison showing the spectra of a representative sample (upper, "bag") and the class-averaged spectra (lower, Class 0). **e**, Averaged magnitude of the spectral FFT (semi-log scale). **f**, Distribution of spectral information entropy across fibers.

Representative reconstructions from a model trained on A1 (Fig. 4c) corroborate these quantitative trends (additional examples provided in Fig. S4 in *Supplementary Note 4*). When applied to unseen data from A2, B2, and B3, the network recovers images that are visually faithful to the ground truth, preserving major semantic contours. For B1, while the objects remain recognizable, the reconstructions exhibit increased blur and reduced contrast consistent with the lower correlation scores in Fig. 4b and spectral entropy.

It is worth noting that to emulate a realistic, calibration-free operational environment, we did not perform individual normalization on the input spectra. As shown in Fig. 4d (upper panel), the absolute intensity of the spectra varies across fibers (e.g., B1 vs. A1) due to differences in coupling efficiency and exposure settings required to avoid saturation. Despite these intensity fluctuations, the model successfully reconstructed the gray-level information of the images (Fig. 4c). This demonstrates that SSDU primarily exploits the relative distribution of spectral features (*i.e.*, spectral shape) rather than absolute intensity values, rendering it inherently tolerant to intensity fluctuations commonly encountered in practical imaging systems (see *Supplementary Note 4*).

Collectively, these results demonstrate that SHINE maintains high cross-fiber reproducibility. Our findings indicate that success is not strictly bound to identical geometric specifications, but rather to achieving a sufficient level of modal mixing. As evidenced by the contrast between B1 and B3, adequate modal scrambling can be achieved either through sufficient fiber length (as in A1, A2, B2) or increased core diameter (as in B3). Crucially, once this geometric threshold for modal mixing is met (e.g., sufficient length or core diameter, as seen with A1, A2, B2, and B3), the system exhibits remarkable robustness to variations in fiber material or manufacturer. This highlights the potential for practical, calibration-free imaging systems using interchangeable fiber probes within established geometric classes.

## Discussion

Our results establish SHINE as a robust paradigm for imaging through dynamic complex media, fundamentally diverging from traditional spatial speckle-based approaches. While TM and DL methods based on speckle intensity are powerful, they are intrinsically fragile; slight fiber distortions decorrelate the speckle field, invalidating the former calibration and hence demanding recalibration of the fiber TM or updating the neural network. By shifting the information carrier from spatial intensity to the optical spectrum, we exploit a degree of freedom that is naturally invariant to linear mode mixing. However, linear spectral encoding has historically been limited by the difficulty of mapping spatial information to frequency profiles with high specificity. In this study, we resolved this by leveraging the nonlinear optical response, where spatial phase variations dictate the output spectrum via angle-dependent phase-matching conditions. This mechanism effectively bridges the spatial and spectral domains, allowing complex spatial information to be inferred from robust spectral signatures.

Compared to existing linear spectral encoding strategies, such as those using scattering media[27], Fabry-Perot etalons[26], or metasurfaces[28], which are typically restricted to sparse or binary targets due to the strict trade-offs between spectral orthogonality and pixel resolution, our SHINE approach leverages the synergy of deterministic nonlinear physics and deep learning to reconstruct complex grayscale images. Crucially, the physical nature of SHINE offers a decisive stability advantage over the rigorous alignment required by linear encoders. While scattering-based methods operate as complex disordered media dependent on fragile positional interference, and structured encoders rely on complex micro-resonances, the bulk phase-matching mechanism of SHINE is intrinsically immune to transverse (x-y) displacements under far-field or Fourier-plane illumination conditions in a homogeneous, transparent crystal. Furthermore, even in the presence of axial ($z$) shifts or rotations, the system does not require a complete re-characterization of the mapping; instead, performance can be maintained by a straightforward adjustment of the phase-matching angle, positioning SHINE as a resilient candidate for robust imaging through dynamic multimode fibers.

Despite these advances, several challenges remain. A primary constraint is the reconstruction of high-frequency details, reflected in the observation that PCC scores generally exceed SSIM metrics. This discrepancy arises from the angular selectivity of the SHG process, which acts as a "soft" low-pass filter: high spatial frequencies correspond to large off-axis propagation angles, which experience lower conversion efficiency and are thus underrepresented in the output spectrum.

Additionally, the spectral bandwidth of the nonlinear crystal restricts the total information throughput. While broadband lasers potentially offer thousands of spectral points, conventional bulk crystals support a narrower phase-matching bandwidth. Future improvements could employ engineered materials, such as quasi-phase-matched structures or dispersive nonlinear metasurfaces[40-44], to broaden the spectral acceptance while keeping the spectral diversity to enhance spatial resolution. For instance, dielectric nanoresonators hosting bound states in the continuum (BIC)[43] and thin-film lithium niobate (LiNbO3)[44] metasurfaces have demonstrated encouraging nonlinear enhancement with tailorable angular and spectral responses. Unlike bulk crystals, these platforms allow the phase-matching condition to be engineered in a more complex manner, broadening the spectral acceptance without sacrificing the discrimination power required for encoding. Furthermore, such metasurfaces possess unique ability to provide spatial resolution directly on the device plane, in contrast to bulk crystals that strictly operate in the angular domain. This capability could eliminate the need for bulky focusing optics and avoid requiring the object to be in the far field, thereby significantly reducing the system complexity and footprint.

Looking forward, practical deployment will also require addressing system integration and biosafety. Building on the potential of such integrated nanophotonic interfaces, the SHINE paradigm is inherently compatible with coherent reflection-mode imaging, conceptually similar to optical coherence tomography (OCT). In a clinical scenario, the fundamental pulse could be delivered through the fiber core (or a double-clad inner core) to the distal tip. By integrating the nonlinear encoder directly at the fiber facet, the backscattered signal from the tissue undergoes SHINE locally before being collected and transmitted back through the fiber. Because the encoding locks spatial information into the spectrum at the distal end, subsequent modal scrambling during the return propagation does not corrupt the image, analogous to the principle of spectrally encoded endoscopy[45,46].

Regarding safety, the high peak intensities required for efficient SHG pose potential phototoxicity risks for in-vivo applications. Future research will explore materials with higher nonlinearity or resonant enhancement strategies to minimize optical power requirements while preserving spectral fidelity.

## Conclusion

In summary, we have demonstrated a calibration- and feedback-free imaging modality that overcomes the longstanding fragility of MMF imaging systems. By encoding spatial information into nonlinear spectral signatures, SHINE achieves high-fidelity image reconstruction and classification that withstands dynamic bending, twisting, and environmental vibration. Uniquely, this method decouples the encoding mechanism from the transmission medium, enabling a "plug-and-play" capability where a model trained on a single fiber generalizes to entirely distinct, previously unseen fibers without calibration or feedback—a feat unattainable by conventional TM and interactive wavefront shaping scheme.

This work represents a fundamental shift in how we approach information transmission through disordered media, moving from correcting complex interference patterns to exploiting invariant physical quantities. By synergizing the physical laws of nonlinear optics with the statistical power of deep learning, SHINE offers a scalable solution for real-time, robust computational imaging. This paves the way for next-generation minimally invasive diagnostics, resilient optical communications, and adaptive sensing systems that operate reliably in the demanding dynamic environments.

## Methods

**Experimental Setup**

The spatially modulated beam was generated by a Ti:sapphire oscillator (Vitara-T, Coherent). The laser operated at a central wavelength of 810 nm, with a repetition rate of 80 MHz, a pulse duration of ~20 fs, and a spectral bandwidth of ~125 nm. The average power was ~300 mW, corresponding to a pulse energy of approximately 3.75 nJ. The beam was then focused by an achromatic doublet lens ($f$=50 mm) into a 200 μm-thick BBO crystal (7 × 7 × 0.2 mm, $\theta$ = 29.2°, $\varphi$ = 0°, CASTECH) to induce second-harmonic generation (SHG). The 200 μm thickness was optimized as a compromise between signal strength and angular acceptance bandwidth, while the $\theta$=29.2° cut was chosen to maximize phase-matching efficiency for the 810 nm fundamental wavelength at normal incidence. Following interaction, a second lens (L2) collimated the beam, and a short-pass filter was employed to isolate the SHG signal from the residual fundamental light. Prior to fiber coupling, the SHG signal was attenuated using a neutral density filter to avoid detector saturation. The filtered signal was then coupled into a MMF using a bi-convex lens. The MMF was coiled into multiple loops and placed on the optical table. Its proximal end was secured by a magnetic base to collect the SHG signal, while the distal end was connected to the spectrometer.

Data acquisition was performed under two distinct conditions to evaluate system robustness. In the static configuration, the MMF remained stationary on the optical table. In the dynamic configuration, the coiled section of the fiber was manually manipulated to simulate a complex, stochastic, and irreproducible environment. This involved applying random mechanical perturbations—including bending, twisting, and vibration. The output spectrum was recorded using a fiber-coupled spectrometer (HR4PRO-VIS-NIR, Ocean Optics, integration time: 12 ms) with a spectral resolution of 0.185 nm/pixel over 390.5–437.7 nm. For each measurement, the spectrum was sampled to 256 points to serve as the input vector for neural network. With the present acquisition settings, frame rates on the order of 10 Hz were readily achievable, which was primarily limited by the spectrometer readout rather than the nonlinear encoding process itself.

**SHG Spectrum Decoding U-net (SSDU) architecture**

We developed a custom convolutional neural network, SSDU, to simultaneously reconstruct spatial images and classify objects based solely on spectral inputs. The SSDU features an encoder-decoder backbone incorporating residual convolutional blocks, designed to produce dual outputs: a 28 × 28-pixel reconstructed image and a 10-class categorical prediction. The classification branch utilizes global max pooling followed by a fully connected layer with softmax activation. The reconstruction branch employs transposed convolutional layers with skip connections to recover fine spatial features from the latent representation. To enhance morphological fidelity, semantic context from the classification prediction is fused into the reconstruction pathway via feature concatenation, providing high-level structural priors to the decoding process.

**Network training and implementation**

The network was implemented in TensorFlow 2.6 and trained on a workstation equipped with an NVIDIA 3090 Ti GPU. Training was conducted for 50 epochs with a batch size of 32. We utilized the Adam optimizer[47] with an initial learning rate of $1 \times 10^{-3}$, subject to an exponential decay rate of 0.92 per epoch. Binary cross-entropy loss was utilized for both the pixel-wise image reconstruction and the classification tasks. Classification accuracy was evaluated by comparing the predicted class (index by maximum probability) against the ground-truth label.

**Data Availability**

All relevant data are available from the corresponding author upon request

**Code Availability**

All relevant code is available from the corresponding author upon request

**Acknowledgments**

This work was supported by Hong Kong Research Grant Council (15125724, 15217721, C7074-21GF), Hong Kong Innovation and Technology Commission (MHP/206/24), National Natural Science Foundation of China (82330061, 81930048), Shenzhen Science and Technology Innovation Commission (JCYJ20220818100202005), Guangdong Science and Technology Commission (2019BT02X105),Guangzhou Science and Technology Projects (2025B01J3013, 2025B03J0097), and Hong Kong Polytechnic University (P0045762, P0049101, P0048314, P0059222, P0054249). Z. C. acknowledges support by the National Natural Science Foundation of China

**Author contributions**

Z.W., and P. L. conceived the idea; Z.W. and H.L. conducted experiments. Z.W., H.L. S.G., Z.C., and P.L. analyzed the data and results and improved the algorithm performance with assistance from S.L., J.C., T.Z., and J.Y.; Z.W., H.L., Z.Y., S.G., Z.C., and P.L. wrote and revised the manuscript; Z.W., H.L., Z.C. and P.L. prepared the Supplementary Materials; Z.Y., J.P., Z.C., S.G., and P.L. supervised the project; Z.W., H.L., Z.Y., J.P., Z.C., S.G., and P.L. proofread and optimized the manuscript.

**Competing Interests Statement**

The authors declare no competing financial interests.

**Additional information**

**Supplementary information**

**Supplementary video**